\def\Statusstring{   }

\documentclass[12pt,letter]{article}

\usepackage{amsmath, amsthm, amssymb}
\usepackage{amssymb,epsfig}

\usepackage{hyperref}
\hypersetup{bookmarks=true}

\makeatletter

\gdef\@punct{.\ \ }  
\def\@sect#1#2#3#4#5#6[#7]#8{%
  \ifnum #2>\c@secnumdepth
     \def\@svsec{}
  \else
     \refstepcounter{#1}\edef\@svsec{%
     \ifnum #2>0{{\csname the#1\endcsname}}.\fi%
    \hskip .5em}
  \fi
  \@tempskipa #5\relax
  \ifdim \@tempskipa>\z@
     \begingroup #6\relax
       \@hangfrom{\hskip #3\relax\@svsec}{\interlinepenalty \@M #8\par}
     \endgroup
     \csname #1mark\endcsname{#7}
     \addcontentsline{toc}{#1}{\ifnum #2>\c@secnumdepth\else
          \protect\numberline{\csname the#1\endcsname}\fi#7}
  \else
     \def\@svsechd{#6\hskip #3\@svsec #8\@punct\csname
#1mark\endcsname{#7}
     \addcontentsline{toc}{#1}{\ifnum #2>\c@secnumdepth \else
          \protect\numberline{\csname the#1\endcsname}\fi#7}}
  \fi
  \@xsect{#5}}

\def\@ssect#1#2#3#4#5{\@tempskipa #3\relax
  \ifdim \@tempskipa>\z@
     \begingroup #4\@hangfrom{\hskip #1}{\interlinepenalty \@M
#5\par}\endgroup
  \else \def\@svsechd{#4\hskip #1\relax #5\@punct}\fi
  \@xsect{#3}}

\newcommand{\imp}{\mathop{\mathrm{imp}}\nolimits}

\def\qed{\hskip 3pt \hbox{\vrule width4pt depth2pt height6pt}}

\newtheorem{Lemma}{Lemma}
\newtheorem{Example}[Lemma]{Example}
\newtheorem{Theorem}[Lemma]{Theorem}
\newtheorem{Proposition}[Lemma]{Proposition}
\newtheorem{Corollary}[Lemma]{Corollary}
\newtheorem{Definition}[Lemma]{Definition}

\usepackage{calc}
\usepackage{tikz}
\usetikzlibrary{decorations.markings}
\tikzstyle{vertex}=[circle, draw, inner sep=0pt, minimum size=6pt]
\newcommand{\vertex}{\node[vertex]}

\tikzset{->-/.style={decoration={
markings,
mark=at position #1 with {\arrow{>}}},postaction={decorate}}}

\begin{document}

\title{Performance analysis of a distributed algorithm for admission control in wireless networks under the $2$-hop interference model}

\author{Ashwin~Ganesan%
  \thanks{International School of Engineering (INSOFE), Mumbai, Maharashtra, India. 
\texttt{ashwin.ganesan@gmail.com}.}
}

\date{}

\maketitle

\vspace{-8.0cm}
\begin{flushright}
  \texttt{\Statusstring}\\[1cm]
\end{flushright}
\vspace{+5.0cm}

\begin{abstract}
\noindent  
A general open problem in networking is: what are the fundamental limits to the performance that is achievable with some given amount of resources? More specifically, if each node in the network has information about only its $1$-hop neighborhood, then what are the limits to performance?  This problem is considered for wireless networks where each communication link has a minimum bandwidth quality-of-service (QoS) requirement. Links in the same vicinity contend for the shared wireless medium. The conflict graph captures which pairs of links interfere with each other and depends on the MAC protocol. In IEEE 802.11 MAC protocol-based networks, when communication between nodes $i$ and $j$ takes place, the neighbors of both $i$ and $j$ remain silent.  This model of interference is called the $2$-hop interference model because the distance in the network graph between any two links that can be simultaneously active is at least $2$.  In the admission control problem studied in the present paper, the objective is to determine, using only localized information, whether a given set of flow rates is feasible.  

In the present work, a distributed algorithm is proposed for this problem, where each node has information only about its $1$-hop neighborhood.  The worst-case performance of the distributed algorithm, i.e. the largest factor by which the performance of this distributed algorithm is away from that of an optimal, centralized algorithm, is analyzed.  Lower and upper bounds on the suboptimality of the distributed algorithm are obtained, and both bounds are shown to be tight.  The exact worst-case performance is obtained for some ring topologies. While distance-$d$ distributed algorithms have been analyzed for the $1$-hop interference model, an open problem in the literature is to extend these results to the $K$-hop interference model, and the present work initiates the generalization to the $K$-hop interference model. 
\end{abstract}

\newpage
\tableofcontents


\section{Introduction}

Real-time applications such as voice over IP (VoIP) and video conferencing require that the transmitted data be received without much delay in order for the data to be useful.  Such applications have fixed requirements independent of the network performance, and so such applications are called inelastic.  The present work considers inelastic applications whose quality-of-service (QoS) requirements are specified in terms of the minimum bandwidth required over single-hop wireless links.  Links in the same vicinity contend for the shared wireless medium, and this interference is modeled by a conflict graph \cite{Jain:Padhye:etal:03}.  Given a conflict graph and the quality-of-service requirements, the admission control problem is to determine whether a given set of flow rates is feasible.  If an application's demand can be satisfied without disrupting the service promised to previously admitted flows, then the new flow is admitted; otherwise it is denied admission.  For an introduction to the flow admission control problem, the reader is referred to \cite{Gerla:Kleinrock:1980} \cite{Bertsekas:Gallager:1992}.  

Distributed algorithms which rely only on localized information scale with the size of the network.  Communicating information from all nodes in the network to a centralized scheduler incurs communication cost, which reduces the battery life of the nodes.  Even if global information is available at a centralized node, solving the problem optimally can be computationally expensive.  In many practical situations, communication overhead is considered to be much more expensive than local computation. In the admission control problem considered in the present work, the globally optimal solution requires computing the fractional chromatic number of a weighted graph.
There is a polynomial time transformation based on the ellipsoid method between the maximum independent set problem and the fractional chromatic number problem, which implies that the problem of computing the fractional chromatic number of a graph is NP-hard \cite{Grotschel:Lovasz:Schrijver:1981}.  Thus, in order to reduce communication overhead and processing cost,  it is desired that the admission control problem be solved using only localized information and with low processing complexity.

More formally, consider a wireless network, modeled by a simple, undirected graph $G=(V,L)$, where $V$ is a set of wireless devices (also known as nodes or vertices), and $L$ is a set of communication links (also known as edges or wireless links).  Each communication link is between a pair of nodes that are within communication radius of each other. Nodes in the same vicinity contend for the shared wireless medium, and this interference is modeled by a \emph{conflict graph} $G_c = (L, L')$. The vertex set of $G_c$ is the set $L$ of links of the network graph $G$, and two links $\ell_i, \ell_j \in L$ are adjacent vertices in $G_c$ if and only if they interfere with each other and so cannot be simultaneously active.  The conflict graph $G_c$ is determined by the MAC protocol or some other assumption.  An independent set in a graph is a subset of vertices in the graph that are pairwise nonadjacent.  Observe that a set of wireless links in the network can be simultaneously active iff the corresponding subset of vertices is an independent set in the conflict graph.  In the sequel, $G$ denotes the network graph and $G_c$ denotes the conflict graph.   

Suppose the QoS requirement for link $\ell$ is a bandwidth of $f(\ell)$ b/s.  If the total bandwidth of the shared wireless medium is $C$ b/s, then the demand of link $\ell \in L$ can be equivalently expressed by  $\tau(\ell) := f(\ell) / C$, the fraction of each unit of time that link $\ell$ demands to be active.  The admission control problem is: given the conflict graph $G_c$ and link demand vector $\tau = (\tau(\ell): \ell \in L)$, determine whether $\tau$ is feasible.  Sufficient conditions for admission control which can be implemented in a distributed manner include the row constraints \cite{Hamdaoui:Ramanathan:03}  
 \cite{Hamdaoui:Ramanathan:05} \cite{Gupta:Musacchio:Walrand:07}, the degree condition \cite{Hamdaoui:Ramanathan:05}, the mixed condition \cite{Hamdaoui:Ramanathan:05}, and the scaled clique constraints \cite{Gupta:Musacchio:Walrand:07}.  The worst-case performance of various sufficient conditions has been analyzed in the literature \cite{Gupta:Musacchio:Walrand:07} \cite{Chaporkar:Kar:Luo:Sarkar:2008} \cite{Ganesan:2008} \cite{Ganesan:2009} \cite{Ganesan:2010} \cite{Ganesan:WN:2014}; these sufficient conditions and their performance guarantees are applicable for arbitrary conflict graphs.
 
 In many applications, the conflict graphs that arise have additional structure which can be exploited to give distributed algorithms that can be implemented efficiently and with close-to-optimal performance.  For instance, under the $1$-hop interference model, two links in the network graph interfere with each other whenever they are incident to a common node. The $1$-hop interference model is also referred to in the literature as the \emph{primary interference model} or the \emph{node-exclusive interference model}. Applications where this interference model arises include Bluetooth networks, where one channel in a piconet is shared by all the links between a master and its slaves \cite{Haartsen:1998}, and CDMA systems when each node is equipped with only one transceiver \cite{Hajek:1984} \cite{Hajek:Sasaki:1988}. For the $1$-hop interference model, the problem of finding a maximum independent set in the conflict graph is equivalent to finding a maximum matching in the network graph, which is solvable in polynomial time \cite{Lovasz:Plummer:1986}.  It has also been shown that the admission control and scheduling problems are  solvable in polynomial time \cite{Hajek:Sasaki:1988}.  For the $1$-hop interference model,  the performance of distance-$d$ distributed algorithms, where each node has information about only its $d$-hop neighborhood, has been analyzed, and the tradeoffs between performance and complexity and between performance and level of decentralization have been characterized \cite{Ganesan:ToN:2020}.  

Given a network graph $G=(V,L)$, the distance in $G$ between links $e$ and $f$ is defined to be the minimum distance between their endpoints.  In other words, if link $e$ is between nodes $u$ and $v$, and link $f$ is between nodes $x$ and $y$, then the distance between links $e$ and $f$ is defined to be the minimum of the four values $d_G(u,x)$, $d_G(u,y)$, $d_G(v,x)$ and $d_G(v,y)$, where $d_G(a,b)$ denotes the distance in $G$ between vertices $a$ and $b$.  For example, the distance between two links that share a common endpoint is $0$, and the distance between two links that are separated by a single edge is $1$.  

Under the \emph{$K$-hop interference model}, two links in $G$ are said to interfere with each other if and only if the distance in $G$ between the two links is less than $K$. Thus, under the $K$-hop interference model, the conflict graph $G_c$ is defined as follows: two links $\ell_i, \ell_j \in L$ are adjacent vertices in $G_c$ iff the distance in $G$ between $\ell_i$ and $\ell_j$ is less than $K$. The conflict graph constructed in this manner from $G$ is denoted $L_K(G)$.  In the special case $K=1$, the conflict graph $L_1(G)$ is sometimes also denoted $L(G)$ and is called the line graph of $G$.  A subset $F \subseteq L$ of edges in a graph $G=(V,L)$ is a \emph{$\delta$-separated matching} $(\delta \ge 1)$ if the distance between any two (distinct) edges in $F$ is at least $\delta$. A $1$-separated matching is usually just called a matching, and a $2$-separated matching is also called a strong matching (or induced matching). Under the $K$-hop interference model, a set $F$ of links of $G$ can be simultaneously active iff $F$ is a $K$-separated matching.  The focus of the present paper is on the $2$-hop interference model, and so in the rest of this paper it is assumed that $K=2$ and $G_c = L_2(G)$. 

\subsection{IEEE 802.11 MAC Protocol and the $2$-hop Interference Model}

Consider a wireless network where two stations $A$ and $C$ transmit at the same to station $B$.  It is possible for the physical environment to be such that $A$ and $C$ do not hear each other and yet they interfere at $B$.  This can happen because of a physical obstruction between $A$ and $B$, resulting in the hidden terminal problem, or due to signal fading.  Due to the resulting collision at $B$, the channel is wasted during the entire duration of $A$'s and $C$'s transmissions.  

To reduce collisions, the IEEE 802.11 MAC protocol has the following optional reservation scheme.  Before $A$ transmits data to $B$, $A$ sends a special type of control frame called a Request To Send (RTS) frame. Station $B$ then replies to $A$ with a Clear To Send (CTS) control frame.  Thereafter, $A$ sends the DATA frame, which contains the data that $A$ originally intended to send to $B$, and $B$ responds with a link-layer acknowledgment (ACK) frame.  If $A$ does not receive an acknowledgment within a certain time period, it retransmits after a random backoff time.  After a certain number of retransmissions, the station $A$ gives up and discards the frame. Because the neighbors of $B$ hear the CTS frame, the neighbors of $B$ also remain idle for the duration of the transmission; the duration of this transmission is specified in the control frames.  

More specifically, the control frames contain duration fields which specify the duration of the transmission.  The duration specified in the RTS frame is calculated to be the time required, in microseconds, for SIFS (i.e. a short interframe space, which is the time spacing maintained between transmission of frames), the CTS frame, SIFS, the DATA frame, SIFS, and the ACK frame.  The time duration specified in the CTS frame is just the duration for the remaining transmission, i.e. is the duration specified in the RTS frame minus the duration for both SIFS and CTS \cite[p. 670]{IEEE:STD:2016}.  This is a \emph{virtual} carrier-sense mechanism in the sense that the nodes learn of the medium reservation from the duration field in the control frames.  Based on this duration information, the network allocation vector (NAV) maintains a counter that counts down to zero at a uniform rate.  While the counter is nonzero, the channel is considered busy. Because these control frames are short and  after the RTS/CTS handshake the subsequent DATA and ACK frames do not experience collisions, the RTS/CTS exchange is useful if the control frames are much shorter than the data frames. 

The interference in IEEE 802.11 MAC protocol-based networks can be modeled by the $2$-hop interference model (also called the \emph{secondary interference model}), which states that two links in the network graph can be simultaneously active if and only if the distance in the network graph between these two links is at least $2$.  Equivalently, two links interfere with each other whenever the distance between them is $0$ or $1$.  One measure of the capacity of a wireless network is the maximum number of simultaneous transmissions that are possible (cf. \cite{Balakrishnan:etal:2004}).  Under the $2$-hop interference model, this quantity is the maximum size of a $2$-separated matching in the network graph, or the maximum size of an independent set in the conflict graph $G_c = L_2(G)$.  The problem of computing the maximum size of a $2$-separated matching in a graph is NP-hard  \cite{Stockmeyer:Vazirani:1982} \cite{Cameron:1989}.  

\subsection{System Model and Problem Formulation}

Consider a wireless network, modeled by a simple, undirected graph $G=(V,L)$, where $V$ is a set of nodes and $L$ is a set of wireless links. Each link $\ell \in L$ makes a demand to be active for a fraction $\tau(\ell)$ of each unit of time.  Assume interference is modeled by the $2$-hop interference model.  This means two links $\ell_i, \ell_j$ interfere with each other iff the distance in $G$ between $\ell_i$ and $\ell_j$ is at most $1$.  Equivalently, the set of links that can be simultaneously active is a $2$-separated matching in $G$.  The admission control problem is the following: given the network graph $G = (V,L)$ and link demand vector $\tau = (\tau(\ell): \ell \in L)$, determine whether $\tau$ is feasible.  That is, determine whether each link can be assigned a subset of $[0,1]$ such that the total duration (or measure) assigned to each link $\ell \in L$ is at least $\tau(\ell)$ and such that links which interfere with each other are assigned subsets that are disjoint (except possibly at the endpoints of intervals). This problem can be equivalently formulated on the conflict graph, as described next.

Given the network graph $G=(V,L)$, the conflict graph $G_c = (L, L')$ is defined to be the graph with vertex set $L$, and with $\ell_i, \ell_j \in L$ being adjacent vertices in $G_c$ iff the distance in $G$ between $\ell_i$ and $\ell_j$ is at most $1$.  Thus, $G_c = L_2(G)$, where $L_K(G)$ is as defined above.   An independent set in a graph is a set of vertices that are pairwise nonadjacent. Let $\mathcal{I}(G_c)$ denote the set of all independent sets of the graph $G_c$.  A \emph{schedule} is a map $t: \mathcal{I}(G_c) \rightarrow \mathbb{R}_{\ge 0}$ that assigns to each independent set in $G_c$ a duration of time when all links in the independent set can be active.  A schedule $t$ \emph{satisfies} demand $\tau$ if $\sum_{I \in \mathcal{I}(G_c): \ell \in I} t(I) \ge \tau(\ell)$, for all $\ell \in L$. The \emph{duration} of the schedule $t$ is $\sum_{I \in \mathcal{I}(G_c)} t(I)$.  Let $T^*(\tau)$ denote the minimum duration of a schedule satisfying $\tau$. A link demand vector $\tau$ is said to be \emph{feasible} if there exists a schedule of duration at most $1$ that  satisfies $\tau$, i.e. if $T^*(\tau) \le 1$. The \emph{admission control problem} is to determine, given a conflict graph $G_c$ and link demand vector $\tau$, whether $\tau$ is feasible.  The independent set polytope $P_I$ is defined to be the convex hull of the characteristic  vectors of the independent sets in $G_c$.  Then, $P_I$ is exactly the set of all link demand vectors that are feasible within $1$ unit of time, and a necessary and sufficient condition for $\tau$ to be feasible is that $\tau \in P_I$. Another equivalent formulation in terms of the fractional chromatic number is given next. 

Let $G_c = (L, L')$ be a conflict graph, where $L = \{\ell_1, \ell_2, \ldots, \ell_N\}$, and let $\mathcal{I}(G_c) = \{I_1, \ldots, I_K\}$ be  the set  of all maximal independent sets of the conflict graph.  Define the $N \times K$ vertex-independent set incidence matrix $B=[b_{ij}]$ by $b_{ij} = 1$ if $\ell_i \in I_j$, and $b_{ij}=0$ otherwise.  The fractional chromatic number of the weighted graph $(G_c, \tau)$, denoted by $\chi_f(G_c,\tau)$, is the optimal value of the linear program: 
$$\min 1^t t \mbox{ subject to } Bt \ge \tau, t \ge 0.$$  
Then, $\chi_f(G_c,\tau)$ is the minimum duration of a schedule satisfying $\tau$, and the admission control problem is equivalent to determining whether $\chi_f(G_c,\tau) \le 1$.   

Some terminology from graph theory that will be used in the sequel is the following. Let $G$ be a graph with vertex set $V(G)$. The subgraph of $G$ induced by $W \subseteq V(G)$, denoted $G[W]$, is the graph with vertex set $W$, and with $uv$ being an edge in $G[W]$ whenever $uv$ is an edge in $G$ and both endpoints $u,v$ belong to $W$.   A clique in a graph is a subset of vertices that are pairwise adjacent.  An induced cycle in $G$ is a cycle that is an induced subgraph. A graph is said to be \emph{chordal} (or triangulated) if it does not contain an induced cycle of length at least $4$.  
It may be assumed in the analysis that follows that the conflict graph $G_c$ is a connected graph, because otherwise the same analysis can be carried out separately on each connected component of the conflict graph.    For any terminology on graph theory not explicitly recalled here, the reader is referred to \cite{Bollobas:1998} \cite{Scheinerman:Ullman:2013}.


\subsection{Summary of Results}
The main contributions of this paper are as follows.

\begin{enumerate}
 \item \emph{A distributed algorithm.}  A fundamental open problem in networking is: what are the limits to the performance that is achievable with some given amount of resources?  More specifically, what is the theoretical best performance that is achievable if each node has information about only its $d$-hop neighborhood.  In \cite{Ganesan:ToN:2020}, this problem was addressed for the $1$-hop interference model, and an open problem in the literature mentioned in \cite[p. 194]{Ganesan:ToN:2020} is to generalize these results from the $1$-hop interference model to the $K$-hop interference model.  The present work initiates this study for the $2$-hop interference model, which arises in IEEE 802.11 MAC protocol-based networks.   A distance-$1$ distributed algorithm is proposed for wireless networks under the $2$-hop interference model. A sufficient condition for admission control is given (Theorem~\ref{thm:suff:cond:2hop:dist1}).
 
 \item \emph{Worst-case performance analysis.}  The worst-case performance of the distance-$1$ distributed algorithm is analyzed.  This quantity is the largest factor by which the network overestimates the resource requirements, and hence is an important parameter to investigate. Lower bounds and upper bounds are obtained - see Corollary~\ref{cor:lowerbound} and Theorem~\ref{thm:suff:cond:2hop:dist1}, respectively. Both bounds are shown to be tight.  The upper bound on the suboptimality of the distributed algorithm gives a worst-case performance guarantee for the distributed algorithm.  The exact worst-case performance is obtained for a class of ring topologies.  
 
 \item \emph{New graph invariants.}  Three new graph invariants are introduced in this paper because they arise in the performance analysis.  Besides the main parameter $\beta(G)$ which characterizes the worst-case performance of the distributed algorithm, two other graph invariants defined in this paper are $\nu(G)$, the maximum size of an exactly-$1$-separated matching, and $\lambda(G)$, the minimum number of $1$-hop neighborhoods in $G$ needed to cover a clique in $L_2(G)$.   These graph invariants were shown to arise in this practical context, and a further study of their properties, complexity, and approximation algorithms for computing them would also be of independent interest.
 
\end{enumerate}

The rest of this paper is organized as follows. In Section~\ref{sec:related:work}, the literature relevant to the present work is mentioned.  Section~\ref{sec:distr:algo} gives a distance-$1$ distributed algorithm for admission control for wireless networks under  $2$-hop interference model.  In Section~\ref{sec:perf:analysis}, a worst-case performance analysis of the distributed algorithm is carried out; lower and upper bounds are obtained for the worst-case performance, and both bounds are shown to be tight.  In particular, the upper bound in Section~\ref{sec:ub} gives a sufficient condition for admission control and a performance guarantee for the distributed algorithm.  The exact worst-case performance is obtained for certain ring networks \ref{sec:ring}.  Finally, Section~\ref{sec:concluding:remarks} contains concluding remarks. 

\section{Related Work} \label{sec:related:work}

The problem of estimating global parameters from local information is a general research area \cite{Linial:1993}. For the problem of the design and analysis of distributed algorithms for admission control for wireless networks under the $K$-hop interference model, the results go back to at least the work of Shannon \cite{Shannon:1949}, whose upper bound on the chromatic index of multigraphs  was used in \cite{Kodialam:Nandagopal:03}
\cite{Kodialam:Nandagopal:05} \cite[p. 1331]{Ganesan:WN:2014} 
to obtain a distance-$0$ distributed algorithm for the $1$-hop interference model.  These results were extended in \cite{Ganesan:2008} \cite{Ganesan:2009} \cite[Theorem 14]{Ganesan:WN:2014} to give a distance-$1$ distributed algorithm for the $1$-hop interference model, using the theory of graph imperfection.  These results were further generalized in \cite{Ganesan:ToN:2020} \cite{Ganesan:WiSPNET:2020}
 to distance-$d$ distributed algorithms for the $1$-hop interference model, for arbitrary $d$.   In a distance-$d$ distributed algorithm, each node has information only about its $d$-hop neighborhood; the tradeoffs between the level of decentralization and performance and between complexity and performance were quantified in \cite{Ganesan:ToN:2020} for the $1$-hop interference model.    

The capacity of a wireless network is related to the maximum number of simultaneous transmissions that can take place.  Under the $K$-hop interference model, this problem is equivalent to computing the maximize size of a $K$-separated matching in the network graph, which can be solved in polynomial time if $K=1$ \cite{Lovasz:Plummer:1986}, but is NP-hard if $K \ge 2$, even for bipartite graphs  
\cite{Stockmeyer:Vazirani:1982} \cite{Cameron:1989}.  For the $2$-hop interference model, it was shown in \cite{Balakrishnan:etal:2004} that for certain networks that arise in practice, the problem can be approximated efficiently.  Greedy algorithms for approximating a maximum weight $2$-separated matching are given in \cite{Lin:etal:2018}. 

Greedy, distributed scheduling algorithms, often referred to in the literature as maximal scheduling, and their worst-case performance, have been investigated by many researchers \cite{Hamdaoui:Ramanathan:05} \cite{Sharma:Mazumdar:Shroff:2006}
\cite{Sharma:Shroff:Mazumdar:2006} \cite{Wu:Srikant:2006} 
\cite{Gupta:Musacchio:Walrand:07} \cite{Ganesan:WN:2014} \cite{Kose:Medard:PIMRC:2017}.  The worst-case performance of these distributed algorithms is characterized by the induced star number (or interference degree) $\sigma(G_c)$ of the conflict graph \cite{Chaporkar:Kar:Luo:Sarkar:2008} \cite{Ganesan:2008} \cite{Ganesan:2009} \cite{Ganesan:2010}.  The induced star number of a graph is the maximum number of leaf vertices in the largest induced star subgraph of the graph \cite{Ganesan:2010}.  Under the $1$-hop interference model, the conflict graph is a line graph, and the induced star number of a line graph is at most $2$. This implies that there exist distributed algorithms for admission control (and also for scheduling) for the $1$-hop interference model which are a factor of at most $2$ away from optimal.  However, for the $K$-hop interference model ($K \ge 2$), the induced star number of the conflict graph is not bounded from above by a constant, and so it is possible for the performance of distributed algorithms to be arbitrarily far away from optimal. An open problem is to design efficient distributed algorithms with performance guarantees for the $K$-hop interference model.

Under additional assumptions, the conflict graph has an induced star number which is bounded from above by a constant that is independent of the size of the network graph. For example, for a certain geometric, unit disk graph model called the bidirectional equal power model, the induced star number of the conflict graph is at most $8$ and this bound is tight \cite{Chaporkar:Kar:Luo:Sarkar:2008}.  In \cite{Joo:etal:2010}, it is shown that if $G_c$ is the conflict graph under the geometric $K$-hop interference model, then $\sigma(G_c)$ is at most $49$.  The ``geometric'' $K$-hop interference model is different from the graph-theoretic model studied in the present paper; in the ``geometric'' model, two nodes in the network are joined by an edge iff the Euclidean distance between them is at most $1$, and two links are adjacent vertices in the conflict graph iff the Euclidean distance between the links is at most $K$.  The conflict graphs arising under these geometric assumptions generally have different properties than the conflict graphs constructed using the graph-theoretic $K$-hop interference model studied in the present paper. Recently, it was shown that the induced star number of the conflict graph of line networks under the protocol interference model is at most $3$ and that this bound is tight \cite{Ganesan:ToN:2020}.

\section{A Distributed Algorithm} \label{sec:distr:algo}

In this section, a distributed algorithm for the flow admission control problem in wireless networks under the $2$-hop interference model is given.  This distributed algorithm is a distance-$1$ distributed algorithm in the sense that each node uses information about only its $1$-hop neighborhood to make its decisions.  Distance-$d$ distributed algorithms for the $1$-hop interference model were investigated in \cite{Ganesan:ToN:2020}; the present work initiates the generalization to $K$-hop interference models.   

Recall that given a network graph $G=(V,L)$ and a link demand vector $\tau = (\tau(\ell): \ell \in L)$, the objective is to determine, using only localized information, whether $\tau$ is feasible, assuming the $2$-hop interference model.   Given $G=(V,L)$, let $G_v$ denote the subgraph of $G$ induced by $\{v\} \dot{\cup} \Gamma(v)$, where $\Gamma(v)$ denotes the set of neighbors in $G$ of node $v$.  Thus, $G_v$ is the ball of radius $1$ centered at $v$, and is referred to in the sequel as the $1$-hop neighborhood in $G$ of node $v$.  Let $T^*(G_v, \tau)$ denote the minimum duration of a schedule satisfying the demands of all links in the subgraph $G_v$.  A distributed algorithm for admission control is the following: if $T^*(G_v, \tau)$ is at most some threshold value (to be defined below), then node $v$ would conclude that the demand $\tau$ is feasible. More specifically, let $T_1^*(\tau) := \max_{v \in V} T^*(G_v, \tau)$.  Then, a sufficient condition for $\tau$ to be feasible is that $T_1^*(\tau)$ be at most some threshold  (see Theorem~\ref{thm:suff:cond:2hop:dist1} below for details).  Note that the minimum durations $T^*(G_v,\tau)$ and $T_1^*(\tau)$ are computed assuming the $2$-hop interference model.   In the present paper, the results are presented in the form of sufficient conditions for admission control (cf. Theorem~\ref{thm:suff:cond:2hop:dist1}), as is commonly done in the literature. These results can be converted to distributed algorithms that are presented in pseudocode form; for example, see \cite{Peleg:2000} \cite[Algorithm~9]{Ganesan:ToN:2020}.

This distance-$1$ distributed algorithm is not optimal in the sense that it is conservative and can overestimate the resource requirements, thereby sometimes rejecting flows that are feasible.  The worst-case performance of this distributed algorithm is an important metric because it is the largest factor by which the distributed algorithm overestimates the resource requirements, when compared to an optimal, centralized algorithm.  A formal definition and analysis of this worst-case performance is given in the next section.  

\section{Performance Analysis} \label{sec:perf:analysis}

In this section, the performance of the distributed algorithm given in the previous section is analyzed.  

\subsection{Preliminaries}

Given a network graph $G=(V,L)$ and link demand vector $\tau = (\tau(\ell): \ell \in L)$, the distance-$1$ distributed algorithm described above computes $T_1^*(\tau) = \max_{v \in V} T^*(G_v, \tau)$, where $G_v$ is the $1$-hop neighborhood of $G$ centered at $v$, and $T^*(G_v,\tau)$ denotes the minimum duration of a schedule satisfying the demands of all links in the induced subgraph $G_v$. The actual resource requirements for satisfying $\tau$, as determined by an optimal, centralized scheduler, is $T^*(\tau) := T^*(G, \tau)$, and takes into account the demands of all links in the entire network graph $G$.  Clearly, $T_1^*(\tau) \le T^*(\tau)$.  The worst-case performance (approximation ratio) of this distributed algorithm is the largest factor by which the estimate $T_1^*(\tau)$ can be away from the actual value $T^*(\tau)$, and is defined by $\sup_\tau \frac{T^*(\tau)}{T_1^*(\tau)}$.  This graph invariant is defined formally next.

\begin{Definition}
Let $G=(V,L)$ be a network graph.  For a given link demand vector $\tau = (\tau(\ell): \ell \in L)$, let $T^*(\tau)$ denote the minimum duration of a schedule satisfying $\tau$ under the $2$-hop interference model, as determined by an optimal, centralized algorithm.  Let $T_1^*(\tau) = \max_{v \in V}  T^*(G_v, \tau)$ denote the estimate computing by the distributed algorithm, again under the $2$-hop interference model.  Then, the worst-case performance $\beta(G)$ is defined by 
$$ \beta(G) := \sup_{\tau} \frac{T^*(\tau)}{T_1^*(\tau)},$$
where the supremum is taken over all nonzero link demand vectors $\tau$. 
\end{Definition}

\begin{Example}  \upshape 
 Let $G=(V,L)$ be the $6$-cycle network graph consisting of the links $\ell_1, \ell_2, \ldots, \ell_6 \in L$, in that cyclic order.  Suppose $\tau = (\tau(\ell_1), \ldots, \tau(\ell_6)) =$ $(1, 0, 1, 0, 1, 0)$. Then, the $1$-hop neighborhood subgraph $G_v$ centered at any node $v \in V$ will contain exactly two neighboring links of $G$. Hence, $G_v$ is isomorphic to $K_{1,2}$.  The demands of the two links in this subgraph are $0$ and $1$, respectively.  The minimum duration of a schedule satisfying $\tau$ in a $1$-hop neighborhood is $T^*(G_v,\tau) = 1$.  The three links $\{\ell_1,\ell_3,\ell_5\}$ interfere with each other, and so the minimum duration $T^*(\tau)$ of a schedule satisfying the demands of all the links is $3$.  Hence, $\frac{T^*(\tau)}{T_1^*(\tau)} = 3$ and $\beta(G) \ge 3$.  This means that for the $6$-cycle network graph, the distance-$1$ distributed algorithm is a factor of at least $3$ away from optimal.  
 \qed
\end{Example}

A necessary and sufficient condition for $\tau$ to be feasible is that $T^*(\tau) \le 1$; however, $T^*(\tau)$ is generally NP-hard to compute and requires that global information be communicated to a single, centralized node.  A centralized solution does not scale for reasons of both computational complexity and communication overhead.  The local estimate $T_1^*(\tau)$ is a lower bound on the actual resources requirements $T^*(\tau)$, and can be computed both efficiently and using only localized information.  A necessary condition for $\tau$ to be feasible is that $T_1^*(\tau) \le 1$.  By scaling this necessary condition, one obtains a sufficient condition:

\begin{Lemma} \label{lem:suff:beta}
 Let $G=(V,L)$ be a network graph and let $\tau$ be a link demand vector.  A sufficient condition for $\tau$ to be feasible is that $T_1^*(\tau) \le \frac{1}{\beta(G)}$. 
\end{Lemma}

\noindent \emph{Proof:}
Let $\tau$ be any link demand vector and suppose $T_1^*(\tau) \le \frac{1}{\beta(G)}$. It follows from the definition of $\beta(G)$ that $T^*(\tau) \le \beta(G) T_1^*(\tau)$.  Hence, $T^*(\tau) \le \beta(G) \frac{1}{\beta(G)} = 1$ and $\tau$ is feasible. 
\qed

The next result shows that the worst-case performance $\beta(G)$ can be arbitrarily large. 

\begin{Lemma}
 The performance of the distributed algorithm given above can be arbitrarily far away from optimal.
\end{Lemma}

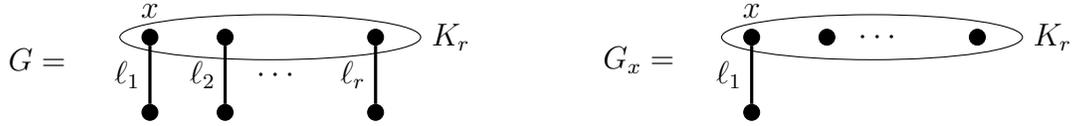
\begin{figure*}
\begin{center}
\begin{tikzpicture}
              
\vertex[fill] (x1) at (0,1) [label=above:$x$] {};
\vertex[fill] (y1) at (0,0) {};
\draw[very thick] (x1) to (y1);
\node at (-0.3, 0.5) {$\ell_1$};

\vertex[fill] (x2) at (1,1) {};
\vertex[fill] (y2) at (1,0) {};
\draw[very thick] (x2) to (y2);
\node at (0.7, 0.5) {$\ell_2$};

\node at (1.7, 0.5) {$\cdots$};

\vertex[fill] (xr) at (3,1) {};
\vertex[fill] (yr) at (3,0) {};
\draw[very thick] (xr) to (yr);
\node at (2.7, 0.5) {$\ell_r$};

\draw (1.6,1) ellipse (2cm and 0.3cm);

\node at (4, 1) {$K_r$};
\node at (-1.5, 0.7) {$G=$};

\vertex[fill] (x1p) at (8,1) [label=above:$x$] {};
\vertex[fill] (y1p) at (8,0) {};
\draw[very thick] (x1p) to (y1p);
\node at (7.7, 0.5) {$\ell_1$};

\vertex[fill] (x2p) at (9,1) {};
\node at (9.7, 1) {$\cdots$};

\vertex[fill] (xrp) at (11,1) {};

\draw (9.6,1) ellipse (2cm and 0.3cm);

\node at (12, 1) {$K_r$};
\node at (6.5, 0.7) {$G_x=$};

\end{tikzpicture}
\caption{(a) A network graph $G$ consisting of a clique $K_r$ and $r$ independent edges. (b) The $1$-hop neighborhood of $G$ centered at node $x$.}
\label{fig:proof:clique:neighborhood}
\end{center}
\end{figure*}

\noindent \emph{Proof:}
In order to show that $\beta(G)$ can be arbitrarily large, it suffices to show that given any positive integer $r$, there exists a network graph $G$ and a link demand vector $\tau$ such that $T^*(\tau) / T_1^*(\tau) \ge r$.  Fix $r \ge 1$. Let $G$ be the network graph shown in Figure~\ref{fig:proof:clique:neighborhood}, consisting of a clique $K_r$ and a set $F$ of $r$ independent edges $\ell_1,\ldots,\ell_r$  incident to the vertices of this clique. Take $\tau(\ell) = 1$ if $\ell \in F$, and $\tau(\ell) = 0$ otherwise.  The maximum value $\max_{v \in V} T^*(G_x, \tau)$ is achieved for some vertex $x$ in the clique.  Let $x$ be any vertex in this clique.  Then, the $1$-hop neighborhood $G_x$ (see Figure~\ref{fig:proof:clique:neighborhood}(b)) contains exactly one edge from $F$, and so $T^*(G_x, \tau) = 1$.  Hence, $T_1^*(\tau) = 1$.  Under the $2$-hop interference model, the links in $F$ are pairwise interfering and must be scheduled at disjoint time slots. Hence, $T^*(\tau) = r$. 
\qed

\subsection{Lower Bounds}

\begin{Definition}
An \emph{exactly-$1$-separated matching} in $G=(V,L)$ is a subset $F \subseteq L$ of edges such that the distance in $G$ between any two (distinct) edges of $F$ is exactly $1$.   Let $\nu(G)$ denote the maximum size of an exactly-$1$-separated matching in $G$. 
\end{Definition}

\begin{Example}  \upshape 
The set $\{\ell_1,\ldots,\ell_r\}$ of links in Figure~\ref{fig:proof:clique:neighborhood} is an exactly-$1$-separated matching.  Let $K_n$ denote the complete graph on $n$ vertices. Then, $\nu(K_n)= \lfloor n/2 \rfloor$ because every matching in the complete graph is also an exactly-$1$-separated matching and the maximum size of a matching in $K_n$ is $\lfloor n/2 \rfloor$.   Let $G$ be the $6$-cycle graph $C_6$. Then the alternating set of links $\{\ell_1, \ell_3, \ell_5\}$ is an exactly-$1$-separated matching in the graph, and $\nu(C_6) = 3$.  Also, it can be verified that $\nu(C_n) = 2$ for all $n \ge 7$. 
\qed
\end{Example}

Under the $2$-hop interference model, the links in an exactly-$1$-separated matching are pairwise interfering and hence must be scheduled at disjoint time slots. It is possible for an exactly-$1$-separated matching, which induces a clique in the conflict graph, to be not entirely contained in any of the $1$-hop neighborhood subgraphs $G_v$ ($v \in V$), and so the quality of the resource estimate computed by the distributed algorithm can be affected by these types of restricted matchings. A lower bound for the worst-case performance $\beta(G)$ is given next.

\begin{Proposition} \label{prop:lb:nu}
 Let $G=(V,L)$ be a network graph.  Then, a lower bound for the worst-case performance $\beta(G)$ is given by 
 $$ \frac{\nu(G)}{\max_{v \in V} \nu(G_v)} \le \beta(G).$$
\end{Proposition}

\noindent \emph{Proof:}
Given a network graph $G=(V,L)$, let $F \subseteq L$ be a maximum exactly-$1$-separated matching of $G$; thus, $F$ is an exactly-$1$-separated matching of maximal cardinality: $|F| = \nu(G)$.  Take the link demand vector $\tau$ to be the characteristic vector of $F$; that is, define $\tau(\ell) =1$ if $\ell \in F$ and $\tau(\ell)=0$ otherwise. Under the $2$-hop interference model, the links in $F$ are pairwise interfering. Hence, $T^*(\tau)=|F|=\nu(G)$. The local estimate $T^*(G_v,\tau)$ computed at node $v$ is equal to the number of links of $F$ that lie in the induced subgraph $G_v$ and is at most $\nu(G_v)$, the maximal size of an exactly-$1$-separated matching in the induced subgraph $G_v$. It follows that 
$$ \frac{\nu(G)}{\max_{v \in V} \nu(G_v)} \le \frac{T^*(\tau)}{T_1^*(\tau)} \le \beta(G).$$
\qed

\begin{Proposition} \label{prop:lb:cycle}
 Given a network graph $G=(V,L)$, let $k$ denote the smallest integer of size at least $2$ such that there exists a cycle of length $4k+2$ in $G$ that is not contained in the $1$-hop neighborhood of any node of $G$. If no such cycle exists, take $k= \infty$. Then, $\frac{2k+1}{2k} \le \beta(G)$. 
\end{Proposition}

\noindent \emph{Proof:} 
Let $C$ be a cycle of length $4k+2$ in the network graph $G$ such that $C$ is not contained entirely in any of the $1$-hop neighborhood subgraphs $G_v$ ($v \in V$).  Let $\ell_1,\ell_2,\ldots,\ell_{4k+2}$ ($k \ge 2$) be the edges (in that order) of the cycle $C$.   Let $M = \{\ell_1,\ell_3, \ldots, \ell_{4k+1} \}$ consist of every other edge of the cycle.  Define the link demand vector $\tau$ to be: $\tau(\ell)=1$ if $\ell \in M$ and $\tau(\ell)=0$ for the remaining links in the network graph.  It suffices to show that $\frac{2k+1}{2k} \le \frac{T^*(\tau)}{T_1^*(\tau)}$. 

Because the link demand is nonzero only for links in $M$, when computing $T^*(\tau)$ it suffices to consider the subgraph of the conflict graph $G_c$ induced by $M$; this subgraph is isomorphic to the odd cycle $C_{2k+1}$.  Hence, $T^*(\tau)$ is the fractional chromatic number of the odd cycle $C_{2k+1}$ ($k\ge2$), which is equal to $\frac{2k+1}{k}$ \cite{Scheinerman:Ullman:2013}.  

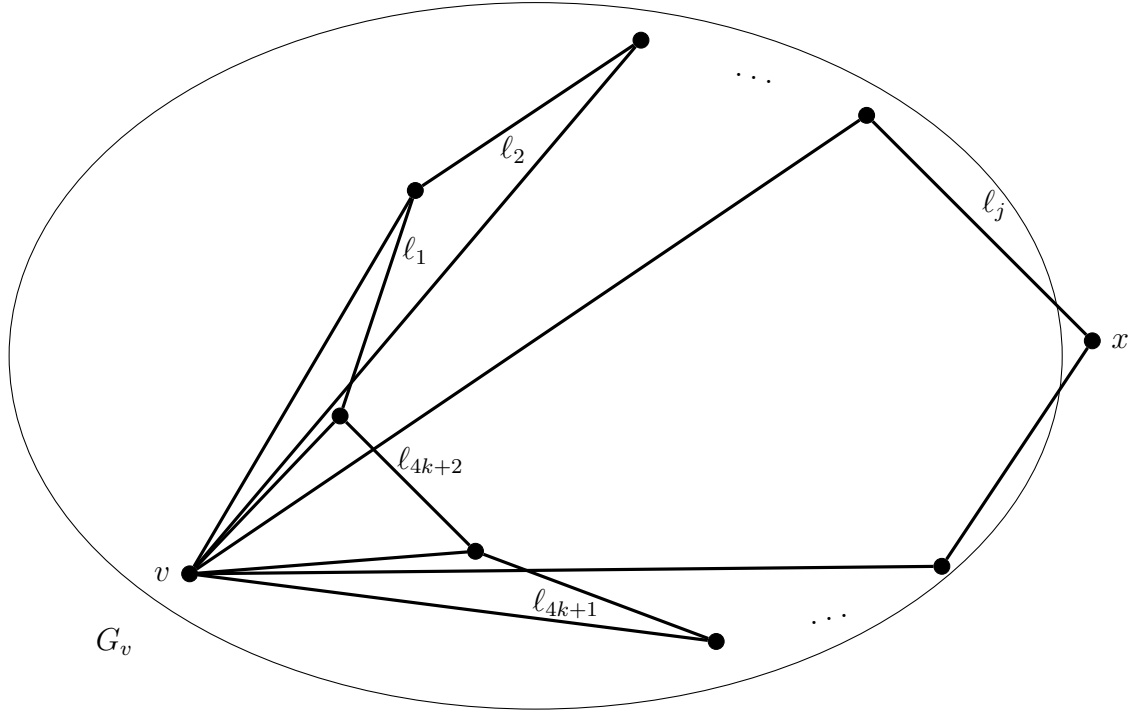
\begin{figure*}
\begin{center}
\begin{tikzpicture}
              
\vertex[fill] (v) at (0,1.9) [label=left:$v$] {};
\vertex[fill] (x1) at (2,4)  {};
\vertex[fill] (x2) at (3,7)  {};
\vertex[fill] (x3) at (6,9)  {};
\vertex[fill] (xj) at (9,8)  {};
\vertex[fill] (x) at (12,5)  [label=right:$x$] {};
\vertex[fill] (xjp2) at (10,2)  {};
\vertex[fill] (x4kp1) at (7,1)  {};
\vertex[fill] (x4kp2) at (3.8,2.2)  {};

\draw[very thick] (v) to (x1);
\draw[very thick] (v) to (x2);
\draw[very thick] (v) to (x3);
\draw[very thick] (v) to (xj);
\draw[very thick] (v) to (xjp2);
\draw[very thick] (v) to (x4kp1);
\draw[very thick] (v) to (x4kp2);

\draw[very thick] (x1) to (x2);
\draw[very thick] (x2) to (x3);
\draw[very thick] (xj) to (x);
\draw[very thick] (x) to (xjp2);
\draw[very thick] (x4kp1) to (x4kp2);
\draw[very thick] (x4kp2) to (x1);

\draw (4.6,4.8) ellipse (7cm and 4.7cm);
\node at (-1, 1) {$G_v$};

\node at (3,6.2) {$\ell_1$};
\node at (4.3,7.55) {$\ell_2$};
\node at (10.7,6.8) {$\ell_j$};
\node at (5,1.5) {$\ell_{4k+1}$};
\node at (3.2,3.4) {$\ell_{4k+2}$};

\node at (7.3,8.55) {$.$};
\node at (7.5,8.5) {$.$};
\node at (7.7,8.45) {$.$};

\node at (8.3,1.25) {$.$};
\node at (8.5,1.3) {$.$};
\node at (8.7,1.35) {$.$};

\end{tikzpicture}
\caption{The case where the $1$-hop neighborhood $G_v$ does not contain some vertex $x$ of the cycle $C_{4k+2}$.}
\label{fig:proof:lb:cycle}
\end{center}
\end{figure*}

Let $v \in V(G)$. It will be shown that the local estimate $T^*(G_v, \tau)$ computed by node $v$, using only information in its $1$-hop neighborhood $G_v$, is at most $2$. By hypothesis, $G_v$ does not contain some vertex, say $x$, of the cycle $C$; see Figure~\ref{fig:proof:lb:cycle}.  Exactly two links from $C$ are incident to $x$, and exactly one of these two links belongs to $M$. This link, call it $\ell_j$, is not contained in $G_v$ because node $x$ does not belong to $G_v$.  It follows that the subgraph of the conflict graph $G_c$ induced by those links which are contained in both $G_v$ and $M$ is bipartite. More simply, the induced subgraph $G_c[M]$ is  an odd cycle and the induced subgraph $G_c[M-\ell_j]$ is bipartite. The fractional chromatic number of a vertex-weighted bipartite graph is the maximum degree of a vertex (cf. \cite{Grotschel:Lovasz:Schrijver:1993}). Hence, $T_1^*(\tau) \le 2$. 
\qed

The two lower bounds for $\beta(G)$ given in Proposition~\ref{prop:lb:nu} and in Proposition~\ref{prop:lb:cycle} can be combined to produce a stronger lower bound:

\begin{Corollary} \label{cor:lowerbound}
 Let $G=(V,L)$ be a network graph. Then, 
 $$ \max\left\{ \frac{\nu(G)}{\max_{v \in V} \nu(G_v)}, \frac{2k+1}{2k} \right\} \le \beta(G),$$
 where $\nu(G)$ is the maximum size of an exactly-$1$-separated matching of $G$, and $k$ is as defined in Proposition~\ref{prop:lb:cycle}. 
\end{Corollary}

\begin{Lemma}
 The lower bound for $\beta(G)$ given in Corollary~\ref{cor:lowerbound} is tight.
\end{Lemma}

\noindent \emph{Proof:}
Take the network graph $G$ to be the complete graph $K_n$.  Then, for each node $v$, the $1$-hop neighborhood $G_v$ is the entire graph. Hence, the exact value of $\beta(G)$ is $1$.  To see that the lower bound also evaluates to $1$, observe that as per the definition of $k$ given above, $k=\infty$.  Because $G_v = G$,  $\nu(G_v) = \nu(G)$ for each node $v$. Hence, $  \frac{\nu(G)}{\max_{v \in V} \nu(G_v)}= 1$. 
\qed

\subsection{Upper Bounds} \label{sec:ub}

A tight upper bound for the worst-case performance $\beta(G)$ will be obtained next.  This gives a bound on the suboptimality of the distributed algorithm, and is used to obtain a sufficient condition for admission control. 

A graph $G_c$ is said to be \emph{perfect} if the chromatic number $\chi(H)$ and clique number $\omega(H)$ are equal for each induced subgraph $H$ of $G_c$.  The imperfection ratio of a graph $G_c$, denoted $\imp(G_c)$, is defined to be $\sup_{\tau \ne 0} \frac{\chi_f(G_c,\tau)}{\omega(G_c,\tau)}$, where $\chi_f(G_c,\tau)$ and $\omega(G_c,\tau)$ denote the fractional chromatic number and clique number of the vertex-weighted graph $(G_c,\tau)$. The imperfection ratio of a graph was studied in \cite{Gerke:McDiarmid:2001}. The imperfection ratio of a graph is at least $1$, and $\imp(G_c) = 1$ iff $G_c$ is perfect. If $H$ is an induced subgraph of $G_c$, then $\imp(H) \le \imp(G_c)$. The odd cycles $C_n$ $(n \ge 5)$ are minimally imperfect graphs, and $\imp(C_n) = \frac{n}{n-1}$. 

A necessary condition for $\tau$ to be feasible is that $\tau(K) \le 1$ for each clique $K$ in the conflict graph. This necessary condition is called the \emph{clique constraints}.  Cliques are local structures, and similar to how the clique number of a graph is a lower bound on the chromatic number, the weighted clique number $\omega(G_c,\tau)$ is a lower bound on the actual resource requirement $\chi_f(G_c,\tau)$.  Their ratio is the factor by which the clique constraints are away from optimal.  This ratio, maximized over all $\tau$, is then the worst-case performance of this distributed algorithm. Hence, the imperfection ratio of a graph characterizes the worst case performance of the clique constraints.

\begin{Definition} \label{def:lambda}
 Let $G=(V,L)$ be a network graph and let $F \subseteq L$ be a set of links. A collection of subgraphs of $G$ is said to \emph{cover} $F$ if the union of the edge set of the subgraphs contains $F$.  The minimum number of $1$-hop neighborhood subgraphs $G_v$ ($v \in V$) needed to cover a set of links in $G$ that are pairwise interfering (under the $2$-hop interference model) is denoted $\lambda(G)$. 
\end{Definition}

\begin{Example}  \upshape 
 Consider the network graph $G=C_{10}$ shown in Figure~\ref{fig:proof:C10}(a).  A $1$-hop neighborhood in the network graph contains three vertices and exactly two links.  Under the $2$-hop interference model, any set of (two or more) pairwise interfering links in the graph is of the form $\{\ell_i,\ell_{i+1}\}$, $\{\ell_i, \ell_{i+1}, \ell_{i+2}\}$ or $\{\ell_i, \ell_{i+2} \}$ for some $i$.  In the first case, the set of two links is contained in a single $1$-hop neighborhood. In the second and third cases, the set is contained in the union of two $1$-hop neighborhoods; for example, observe from Figure~\ref{fig:proof:C10}(a) that $\{\ell_1, \ell_2, \ell_3\}$ is a set of pairwise interfering links and is contained in $G_x \cup G_y$. Hence, $\lambda(C_{10}) = 2$. 
 \qed
\end{Example}

\begin{Theorem} \label{thm:ub:beta:imp:lambda}
Let $G=(V,L)$ be a network graph. Then, the worst-case performance $\beta(G)$ of the distance-$1$ distributed algorithm for the $2$-hop interference model is bounded as
$$ \beta(G) \le \imp(L_2(G)) \lambda(G),$$
where $\imp(L_2(G))$ denotes the imperfection ratio of the conflict graph $L_2(G)$, and $\lambda(G)$ is as per Definition~\ref{def:lambda}. 
\end{Theorem}

\noindent \emph{Proof:}
Let $G=(V,L)$ be a network graph.  Let $G_c=(L, L')$ denote the conflict graph constructed using the $2$-hop interference model.  For a link demand vector $\tau$, let $\omega(G_c,\tau)$ denote the maximum weight of a clique in the weighted graph $(G_c,\tau)$.  

A clique in the conflict graph corresponds to a set $F$ of pairwise interfering links in the network graph $G$.  By definition, $F$ can be covered by $\lambda = \lambda(G)$ or fewer $1$-hop neighborhood subgraphs $G_{v_1}, G_{v_2}, \ldots, G_{v_\lambda}$.  Suppose that the demands of all links in a subgraph $G_{v_i}$ can be satisfied by a schedule of duration at most $1$, for each $i = 1,2,\ldots,\lambda$. Then, by concatenating these schedules one obtains a schedule of duration at most $\lambda$ satisfying the demands of all links in $F$. Hence, $\sup_{\tau: T_1^*(\tau)=1} \omega(G_c,\tau) \le \lambda(G)$. 

Thus,
\begin{align*}
 \beta(G) & = \sup_\tau \frac{T^*(\tau)}{T_1^*(\tau)} \\
 & = \sup_{\tau: T_1^*(\tau)=1} T^*(\tau) \\
 & = \sup_{\tau: T_1^*(\tau)=1} \left(\frac{T^*(\tau)}{\omega(G_c,\tau)}~~ \omega(G_c,\tau) \right) \\
& \le \left( \sup_{\tau: T_1^*(\tau)=1} \frac{T^*(\tau)}{\omega(G_c,\tau)} \right) \left(\sup_{\tau: T_1^*(\tau) = 1} \omega(G_c,\tau) \right)  \\
& \le \left(\sup_\tau \frac{T^*(\tau)}{\omega(G_c,\tau)} \right) \lambda(G) \\
& = \imp(L_2(G)) ~\lambda(G).
\end{align*}
\qed

\begin{Lemma}
 The upper bound given in Theorem~\ref{thm:ub:beta:imp:lambda} is tight.
\end{Lemma}

\noindent \emph{Proof:}
It suffices to show that there exist a network graph $G$ for which the upper bound given in Theorem~\ref{thm:ub:beta:imp:lambda} is exact.  Take $G=C_{10}$.   It will be shown in Section~\ref{sec:ring} that the upper bound evaluates to $2.5$ and that this value is the exact value of $\beta(C_{10})$.
\qed

The upper bound given for $\beta(G)$ can be used to obtain a sufficient condition and a distance-$1$ distributed algorithm for admission control: 

\begin{Theorem} \label{thm:suff:cond:2hop:dist1}
 Let $G=(V,L)$ be a network graph, and let $\tau = (\tau(\ell): \ell \in L)$ be a link demand vector.  Let $T^*(G_v, \tau)$ denote the minimum duration of a schedule satisfying the demands of all links in the $1$-hop neighborhood $G_v$ of node $v$.  Then, under the $2$-hop interference model, a sufficient condition for $\tau$ to be feasible is that $T^*(G_v,\tau) \le \frac{1}{\imp(L_2(G)) \lambda(G)}$, for all $v \in V$.  Here, $\imp(L_2(G))$ denotes the imperfection ratio of the conflict graph $L_2(G)$, and $\lambda(G)$ denotes the minimum number of $1$-hop neighborhood subgraphs in $G$ needed to cover a set of pairwise interfering links of $G$.
\end{Theorem}

\noindent \emph{Proof:}
The assertion follows from Lemma~\ref{lem:suff:beta} and Theorem~\ref{thm:ub:beta:imp:lambda}. 
\qed

\subsection{Ring Topologies} \label{sec:ring}
Recall that $\beta(G)$ characterizes the worst-case performance of the  distance-$1$ distributed algorithm under the $2$-hop interference model.  The exact value of $\beta(G)$ is now obtained for the special case when the network graph $G$ is a cycle graph $C_n$. The case where $n$ is of the form $4k+2$ is considered here. 

\begin{Theorem}
 Suppose the network graph $G$ is the ring topology $C_{4k+2}$ ($k \ge 2$).  Then, the worst-case performance of the distance-$1$ distributed algorithm under the $2$-hop interference model is given by
 $\beta(G) = \frac{2k+1}{k}$. 
\end{Theorem}

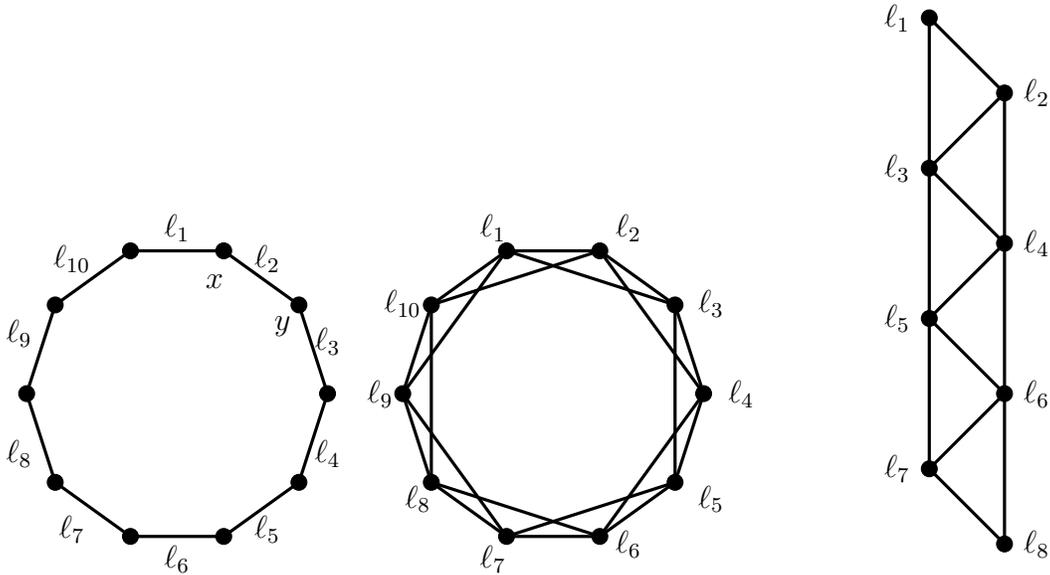
\begin{figure*}
\begin{center}
\begin{tikzpicture}
\vertex[fill] at (2.00,0.00) {};         
\draw[very thick] (2.00,0.00) to (1.62,1.18);           
\vertex[fill] at (1.62,1.18) {};         
\draw[very thick] (1.62,1.18) to (0.62,1.90);           
\vertex[fill] at (0.62,1.90) {};         
\draw[very thick] (0.62,1.90) to (-0.62,1.90);           
\vertex[fill] at (-0.62,1.90) {};         
\draw[very thick] (-0.62,1.90) to (-1.62,1.18);           
\vertex[fill] at (-1.62,1.18) {};         
\draw[very thick] (-1.62,1.18) to (-2.00,0.00);           
\vertex[fill] at (-2.00,0.00) {};         
\draw[very thick] (-2.00,0.00) to (-1.62,-1.18);           
\vertex[fill] at (-1.62,-1.18) {};         
\draw[very thick] (-1.62,-1.18) to (-0.62,-1.90);           
\vertex[fill] at (-0.62,-1.90) {};         
\draw[very thick] (-0.62,-1.90) to (0.62,-1.90);           
\vertex[fill] at (0.62,-1.90) {};         
\draw[very thick] (0.62,-1.90) to (1.62,-1.18);           
\vertex[fill] at (1.62,-1.18) {};         
\draw[very thick] (1.62,-1.18) to (2.00,0.00);           
\vertex[fill] at (7.00,0.00) {};         
\draw[very thick] (7.00,0.00) to (6.62,1.18);           
\draw[very thick] (7.00,0.00) to (5.62,1.90);           
\vertex[fill] at (6.62,1.18) {};         
\draw[very thick] (6.62,1.18) to (5.62,1.90);           
\draw[very thick] (6.62,1.18) to (4.38,1.90);           
\vertex[fill] at (5.62,1.90) {};         
\draw[very thick] (5.62,1.90) to (4.38,1.90);           
\draw[very thick] (5.62,1.90) to (3.38,1.18);           
\vertex[fill] at (4.38,1.90) {};         
\draw[very thick] (4.38,1.90) to (3.38,1.18);           
\draw[very thick] (4.38,1.90) to (3.00,0.00);           
\vertex[fill] at (3.38,1.18) {};         
\draw[very thick] (3.38,1.18) to (3.00,0.00);           
\draw[very thick] (3.38,1.18) to (3.38,-1.18);           
\vertex[fill] at (3.00,0.00) {};         
\draw[very thick] (3.00,0.00) to (3.38,-1.18);           
\draw[very thick] (3.00,0.00) to (4.38,-1.90);           
\vertex[fill] at (3.38,-1.18) {};         
\draw[very thick] (3.38,-1.18) to (4.38,-1.90);           
\draw[very thick] (3.38,-1.18) to (5.62,-1.90);           
\vertex[fill] at (4.38,-1.90) {};         
\draw[very thick] (4.38,-1.90) to (5.62,-1.90);           
\draw[very thick] (4.38,-1.90) to (6.62,-1.18);           
\vertex[fill] at (5.62,-1.90) {};         
\draw[very thick] (5.62,-1.90) to (6.62,-1.18);           
\draw[very thick] (5.62,-1.90) to (7.00,0.00);           
\vertex[fill] at (6.62,-1.18) {};         
\draw[very thick] (6.62,-1.18) to (7.00,0.00);           
\draw[very thick] (6.62,-1.18) to (6.62,1.18);           

\node at (0,2.2) {$\ell_1$};
\node at (0.5, 1.5) {$x$};
\node at (1.2,1.8) {$\ell_2$};
\node at (1.4, 0.9) {$y$};
\node at (2,0.7) {$\ell_3$};
\node at (2,-0.8) {$\ell_4$};
\node at (1.2,-1.8) {$\ell_5$};
\node at (0,-2.2) {$\ell_6$};
\node at (-1.4,-1.8) {$\ell_7$};
\node at (-2.1,-0.8) {$\ell_8$};
\node at (-2.1,0.8) {$\ell_9$};
\node at (-1.4,1.8) {$\ell_{10}$};

\node at (4.2,2.2) {$\ell_1$};
\node at (6,2.2) {$\ell_2$};
\node at (7.1,1.2) {$\ell_3$};
\node at (7.5,0) {$\ell_4$};
\node at (7.1,-1.4) {$\ell_5$};
\node at (6,-2) {$\ell_6$};
\node at (4.2,-2.2) {$\ell_7$};
\node at (3.2,-1.4) {$\ell_8$};
\node at (2.7,0) {$\ell_9$};
\node at (3,1.2) {$\ell_{10}$};

\vertex[fill] at (10.00,5.00) [label=left:$\ell_1$] {};         %
\vertex[fill] at (11.00,4.00) [label=right:$\ell_2$] {};         %
\vertex[fill] at (10.00,3.00) [label=left:$\ell_3$] {};         %
\vertex[fill] at (11.00,2.00) [label=right:$\ell_4$] {};         %
\vertex[fill] at (10.00,1.00) [label=left:$\ell_5$] {};         %
\vertex[fill] at (11.00,0.00) [label=right:$\ell_6$] {};         %
\vertex[fill] at (10.00,-1.00) [label=left:$\ell_7$] {};         %
\vertex[fill] at (11.00,-2.00) [label=right:$\ell_8$] {};         %
\draw[very thick] (10.00,5.00) to (11.00,4.00);           %
\draw[very thick] (11.00,4.00) to (10.00,3.00);           %
\draw[very thick] (10.00,3.00) to (11.00,2.00);           %
\draw[very thick] (11.00,2.00) to (10.00,1.00);           %
\draw[very thick] (10.00,1.00) to (11.00,0.00);           %
\draw[very thick] (11.00,0.00) to (10.00,-1.00);           %
\draw[very thick] (10.00,-1.00) to (11.00,-2.00);           %
\draw[very thick] (10.00,5.00) to (10.00,3.00);           %
\draw[very thick] (10.00,3.00) to (10.00,1.00);           %
\draw[very thick] (10.00,1.00) to (10.00,-1.00);           %
\draw[very thick] (11.00,4.00) to (11.00,2.00);           %
\draw[very thick] (11.00,2.00) to (11.00,0.00);           %
\draw[very thick] (11.00,0.00) to (11.00,-2.00);           %

\end{tikzpicture}
\caption{(a) The network graph $G=C_{10}$. (b) Its conflict graph $G_c$ under the $2$-hop interference model. (c) The subgraph $G_c - \ell_9 - \ell_{10}$, which is chordal.}
\label{fig:proof:C10}
\end{center}
\end{figure*}

\noindent \emph{Proof:}
It will be proved that $\sup_{\tau} \frac{T^*(\tau)}{T_1^*(\tau)} =\frac{2k+1}{k}$.  Let $\ell_1,\ell_2,\ldots,\ell_{4k+2}$ be the links (in that order) of the cycle graph $G = C_{4k+2}$, and let $M = \{\ell_1,\ell_3,\ldots,\ell_{4k+1}\}$. Figure~\ref{fig:proof:C10} shows the network graph $G$ and the conflict graph $G_c = L_2(G)$ for the $k=2$ case; the proof given below holds for arbitrary $k \ge 2$. 

Let $\tau$ be the link demand vector defined by $\tau(\ell)=1$ if $\ell \in M$ and $\tau(\ell)=0$ otherwise.  Under the $2$-hop interference model, the subgraph of the conflict graph induced by $M$ is the odd cycle of length $2k+1$.  Hence, $T^*(\tau)$ is the fractional chromatic number of $C_{2k+1}$ and equals $\frac{2k+1}{k}$.   A $1$-hop neighborhood in $G$ contains exactly two consecutive links of the cycle $G$, and so contains exactly one link of $M$. Hence, $T_1^*(\tau) = 1$, and  $\frac{T^*(\tau)}{T_1^*(\tau)} = \frac{2k+1}{k}$. It follows that $\beta(G) \ge \frac{2k+1}{k}$. 

To prove the opposite inequality, consider the upper bound $\beta(G) \le \imp(L_2(G)) \lambda(G)$ given in Theorem~\ref{thm:ub:beta:imp:lambda}, where $\imp(L_2(G))$ denotes the imperfection ratio of the conflict graph and $\lambda(G)$ denotes the minimum number of $1$-hop neighborhood subgraphs of $G$ needed to cover a (maximal) set of pairwise interfering links.  It suffices to show that $\imp(L_2(G))  = \frac{2k+1}{2k}$ and $\lambda(G) \le 2$.   

A maximal set of pairwise interfering links of $G$ is exactly a set of $3$ consecutive links in the network graph $G$ and hence can be covered by two $1$-hop neighborhood subgraphs of $G$.  For example, observe from Figure~\ref{fig:proof:C10} that $\{\ell_1, \ell_2, \ell_3\}$ is a maximal set of pairwise interfering links of $G$ and is contained in the union of the neighborhood subgraphs $G_x$ and $G_y$. Thus, $\lambda(G) \le 2$.  

It will now be proved that $\imp(G_c) := \sup_\tau \frac{\chi_f(G_c,\tau)}{\omega(G_c,\tau)} = \frac{2k+1}{2k}$, where $G_c$ denotes the conflict graph $L_2(G)$.  Under the $2$-hop interference model,  $\ell_i$ $(i=1,2,\ldots,4k+2)$ is adjacent in the conflict graph to $\ell_{i-2}, \ell_{i-1}, \ell_{i+1}, \ell_{i+2}$. Here, the subscripts are to be interpreted as belonging to one of the congruence classes $1,2,\ldots,4k+2$.  This conflict graph is isomorphic to the Cayley graph $(\mathbb{Z}_{4k+2}, \{ \pm 1, \pm 2\})$ of the cyclic group $\mathbb{Z}_{4k+2}$ with respect to generator set $\{\pm 1, \pm 2\}$.  The generators $\{\pm 2\}$ induce an odd cycle of length $2k+1$ on the vertex set $\{0,2,4,\ldots,4k\}$ in the Cayley graph.  Because the conflict graph $G_c$ contains $C_{2k+1}$ as an induced subgraph, $\imp(G_c) \ge \imp(C_{2k+1}) = \frac{2k+1}{2k}$.  

It remains to be shown that $\imp(G_c) \le \frac{2k+1}{2k}$. Let $\tau$ be any link demand vector. The subgraph $G_c - \ell_{4k+1} - \ell_{4k+2}$ is chordal (see Figure~\ref{fig:proof:C10}(c) for the $k=2$ case) because it does not contain any induced cycles of length at least $4$. Because a chordal graph is perfect \cite{Grotschel:Lovasz:Schrijver:1993}, the fractional chromatic number and clique number of the vertex-weighted graph $(G_c - \ell_{4k+1} - \ell_{4k+2},\tau)$ are equal.  More generally, the weighted graph $(G_c - \ell_i - \ell_{i+1}, \tau)$ can be fractionally colored using $\omega(G_c - \ell_i - \ell_{i+1}, \tau)$ colors, for each $i=1,3,\ldots,4k+1$. Also, $\omega(G_c - \ell_i - \ell_{i+1}) \le \omega(G_c,\tau)$. Putting these colorings together, it is seen that the weighted graph $(G_c, 2k \tau)$ can be fractionally colored using $(2k+1) \omega(G_c,\tau)$ colors.  In other words, $2k \chi_f(G_c,\tau) \le (2k+1) \omega(G_c,\tau)$, as was to be shown.
\qed

\section{Concluding Remarks} \label{sec:concluding:remarks}

A fundamental problem in networking is to investigate the limits to the performance that is achievable with some given amount of resources.  If each node in a wireless network has information about only its $d$-hop neighborhood, then what are the limits to performance?  Distance-$d$ distributed algorithms for admission control have been studied for wireless ad hoc networks under the $1$-hop interference model, and an open problem mentioned in the literature \cite[p. 194]{Ganesan:ToN:2020} is to generalize these  results to $K$-hop interference models. The present work initiates this generalization to the $K$-hop interference model. The $2$-hop interference model arises in IEEE 802.11 MAC protocol-based networks and is the focus of the present work. 

In the present paper, a distance-$1$ distributed algorithm was proposed for this problem and its worst case performance was analyzed.  Upper and lower bounds were obtained and both bounds were shown to be tight.  In particular, the upper bound is used to give a sufficient condition for admission control and also gives a performance guarantee on the distributed algorithm.  Besides the worst-case performance $\beta(G)$, two other new graph invariants were defined in this paper: $\nu(G)$, the maximum size of an exactly-$1$-separated matching, and $\lambda(G)$, the minimum number of $1$-hop neighborhood subgraphs of $G$ needed to cover a clique in $L_2(G)$.  The exact worst case performance of the distance-$1$ distributed algorithm was obtained for certain ring topologies.

These results can be extended in several directions.  First, further properties of the worst-case performance $\beta(G)$ and the two other graph invariants $\nu(G)$ and $\lambda(G)$ defined in this paper can be investigated.  Bounds, complexity analysis, and approximation algorithms for computing them, would also be of independent theoretical interest.   Second, a general research direction is to design and analyze the performance of distance-$d$ distributed algorithms for the $K$-hop interference model, for arbitrary $d \ge 1$, $K \ge 2$, and in particular to analyze the tradeoff between performance and the level of decentralization, and the tradeoff between performance and complexity. 
A third direction is to investigate distributed algorithms for wireless networks for more general combinatorial interference models such as weighted conflict graphs or hypergraphs. Hypergraph interference models were studied in \cite{McEliece:Sivarajan:1994} \cite{Sarkar:Sivarajan:1998}, and continue to be of current interest \cite{Li:Negi:2012} \cite{Halldorsson:Wattenhofer:2019}  \cite{Halldorsson:Tonoyan:2018} \cite{Ganesan:TechRep:2019}.  

 {
\bibliographystyle{myplain}
\bibliography{refs_ag.bib}

}
\end{document}